\documentclass[a4paper,11pt]{article}
\usepackage{jinstpub} 
\usepackage{lineno}

\title{\boldmath Pipeline for performance evaluation of flavour tagging dedicated Graph Neural Network algorithms}

\author[a,b,c]{Greta Brianti}
\author[a,b]{Roberto Iuppa}
\author[c]{Marco Cristoforetti}

\affiliation[a]{University of Trento,\\
   Via Sommarive 14, 38123 Povo, Trento (TN), Italy}
\affiliation[b]{Trento Institute of Fundamental Physics and Applications INFN-TIFPA,\\
Via Sommarive, 14, 38123 Povo, Trento (TN), Italy}
   
\affiliation[c]{Fondazione Bruno Kessler, \\
Via Sommarive 18, 38123 Povo, Trento (TN), Italy}
\emailAdd{greta.brianti@unitn.it}

\abstract{Machine Learning is a rapidly expanding field with a wide range of applications in science. In the field of physics, the Large Hadron Collider, the world's largest particle accelerator, utilizes Neural Networks for various tasks, including flavour tagging. Flavour tagging is the process of identifying the flavour of the hadron that initiates a jet in a collision event, and it is an essential aspect of various Standard Model and Beyond the Standard Model studies. Graph Neural Networks are currently the primary machine-learning tool used for flavour tagging. Here, we present the AUTOGRAPH pipeline, a completely customizable tool designed with a user-friendly interface to provide easy access to the Graph Neural Networks algorithm used for flavour tagging. }

\begin{document}
\maketitle
\flushbottom

\section{Introduction}
\noindent The purpose of flavour tagging is to determine the initial parton of jets, i.e. collimated cone of stable particles arising from fragmentation and the hadronization of a quark after a collision. Unique features of the heavy hadrons (bound states involving bottom or charm quarks) are exploited to identify heavy jets; indeed their decays exhibit distinctive topologies due to their lifetimes. The flavour tagging is particularly important for studying the Standard Model (SM) Higgs boson and the top quark and, additionally, in searches for several Beyond Standard Model (BSM) resonances. The Large Hadron Collider (LHC) at CERN \cite{LHC} involves Machine Learning techniques for online tagging of heavy flavours \cite{algo}. This approach enables the efficient management of large amounts of data, which is crucial for handling the integrated luminosity produced during the last years of Run 2 and expected during Run 3 \cite{run2}. LHC general-purpose experiments, CMS \cite{CMS} and ATLAS \cite{ATLAS}, utilize Graph Neural Networks (GNNs) for flavour tagging. GNNs have a distinct capability of carrying out multi-level inferences. For example, in the ATLAS experiment, the new GN1 tagger \cite{GN1} makes predictions on three levels: the graph classification (i.e. flavour tagging), the node classification (i.e. prediction of the track truth origin), and the edge classification (i.e. whether the tracks in the track-pair belong to a common vertex). Furthermore, GNNs can include the structure of physical events. For example, in the CMS experiment, the ParticleNet tagger \cite{ParticleNet} exploits the particle cloud structures to reconstruct and tag the jets. Finally, GNN algorithms are applied to offline analysis for background-signal classification tasks at colliders \cite{GNN_at_collider}. GNNs are involved in non-collider experiments as well. For instance, IceCube has developed a GNN algorithm for neutrino reconstruction \cite{IceCube}, which improves reconstruction accuracy for low-energy events. It can be inferred that GNNs are a crucial new technology in particle physics, but managing and utilizing them is complex; for this reason, a pipeline that allows easy access to this state-of-the-art technology has been developed. 
\section{The AUTOGRAPH pipeline}
\begin{figure}[htbp!]
    \centering
    \includegraphics[scale=0.32]{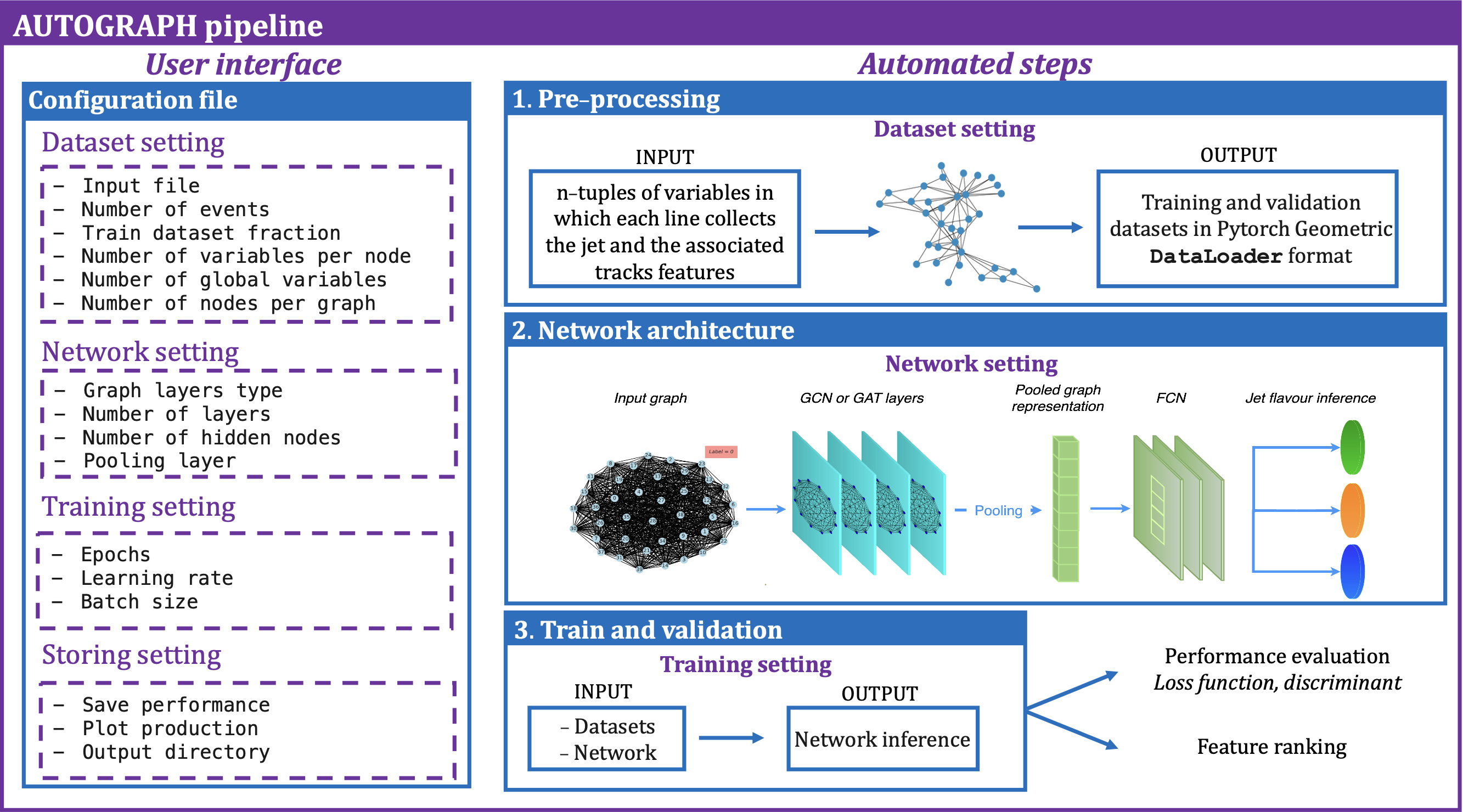}
    \caption{Pictorial pipeline representation. The \textit{user interface} section represents the configuration file wherein the user can choose the setting for the \textit{automated steps}. The latter are divided into three main processes illustrated schematically.}
    \label{fig:pipeline}
\end{figure}

\noindent The AUTOGRAPH pipeline (Automatic Unified Training and Optimization for Graph Recognition and Analysis with Pipeline Handling) is an easy-to-use and customizable architecture designed to train and apply GNNs for flavour tagging. The framework consists of two main components - the user interface and the automated steps, as depicted in Figure \ref{fig:pipeline}. To manage the jet-graph structure, network architecture, and training hyperparameters, the user fills a single configuration file via the provided interface. The pipeline's automated structure consists of Python modules managed by the user interface. To start the training process, the user executes a single Python script.
When it comes to supervised Machine Learning algorithms like the one used in the pipeline, the training process involves adjusting the network parameters (also known as weights) to minimize the loss function. The loss function represents the difference between the network's prediction and the target value associated with the event, typically obtained from the Monte Carlo simulations. The automatic training process consists of three main steps: pre-processing the data, defining the network architecture, and iterating through epochs to update the network weights. Moreover, the network is tested on a validation dataset that is different from the one used for training to assess its performance. 
\subsection{Dataset handling}
\begin{figure}[htbp!]
    \centering
    \includegraphics[scale=0.3]{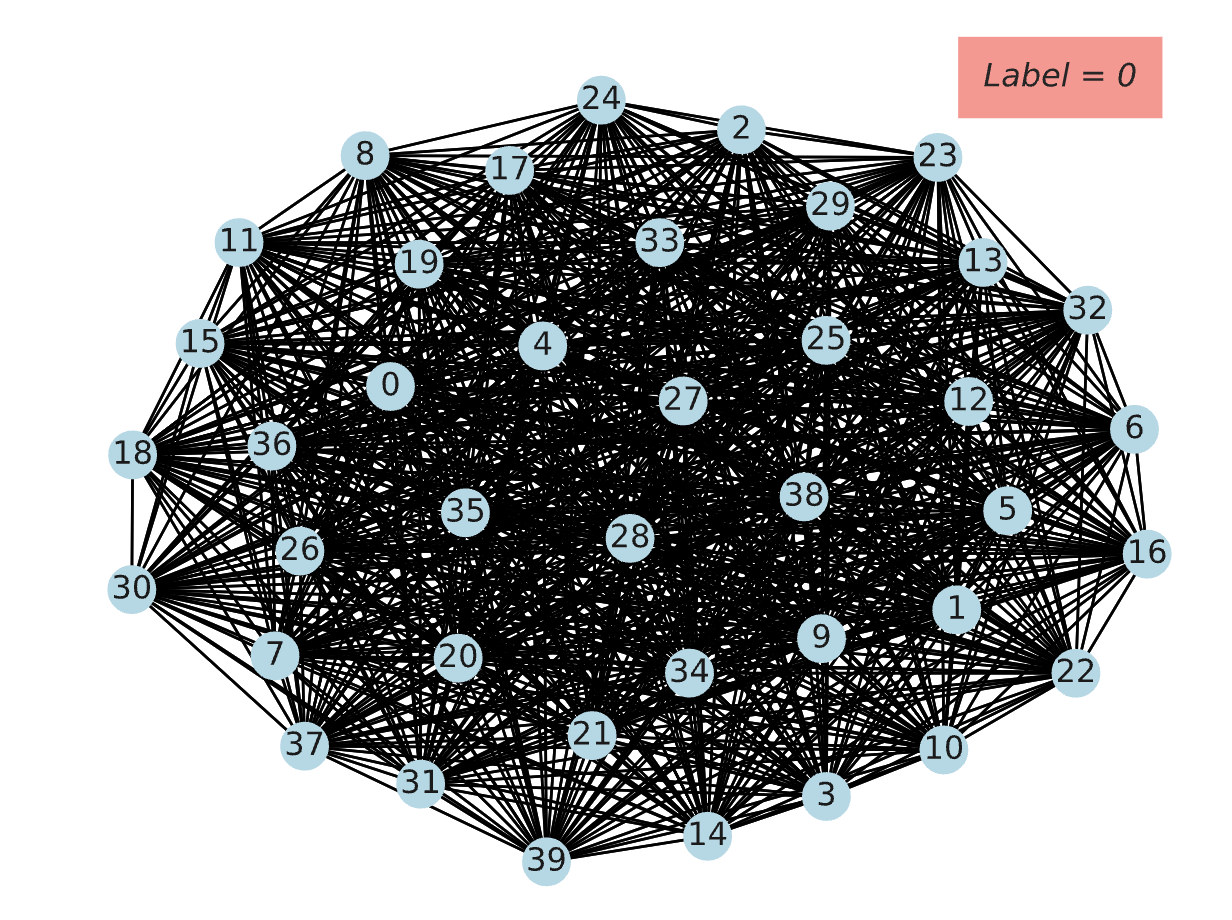}
    \caption{Fully connected graph representation of a jet in the pipeline. The forty nodes are the tracks associated with the jet. The graph is labeled from the Monte Carlo, in this case as 0, corresponding to light jets.}
    \label{fig:grp}
\end{figure}

\noindent The pipeline works on simulated datasets previously pre-processed by the user. It requires an array representation of track-jets with a fixed number of associated tracks. The tracks should be associated with the jets before the pipeline action. In AUTOGRAPH, the dataset undergoes an initial automated step to define the graph representation. The graph in Figure \ref{fig:grp} illustrates a jet, with the tracks linked to the jet represented as nodes in the graph. Each node corresponds to a customizable set of track features the user selects. Finally, once the tracks and their related variables are specified, the user selects the global features associated with the graph structure and links directly with the jet.
Through the configuration file, the user can customize the graph architecture, selecting the number of tracks per jet, the features associated with each track, and the global-jet features. Finally, the resulting graph list is converted into Pytorch Geometric DataLoader format \cite{PyG}. During the conversion, the original dataset is divided into two data sets: the training dataset, dedicated to the network's training, wherein the network parameters are updated at each epoch, and the validation dataset, used to control the network performance. The configuration file allows the user to choose the fraction of data used for training, which is set to 0.7 by default. 

\subsection{Network architecture}
\begin{figure}[htbp!]
    \centering
    \includegraphics[scale=0.09]{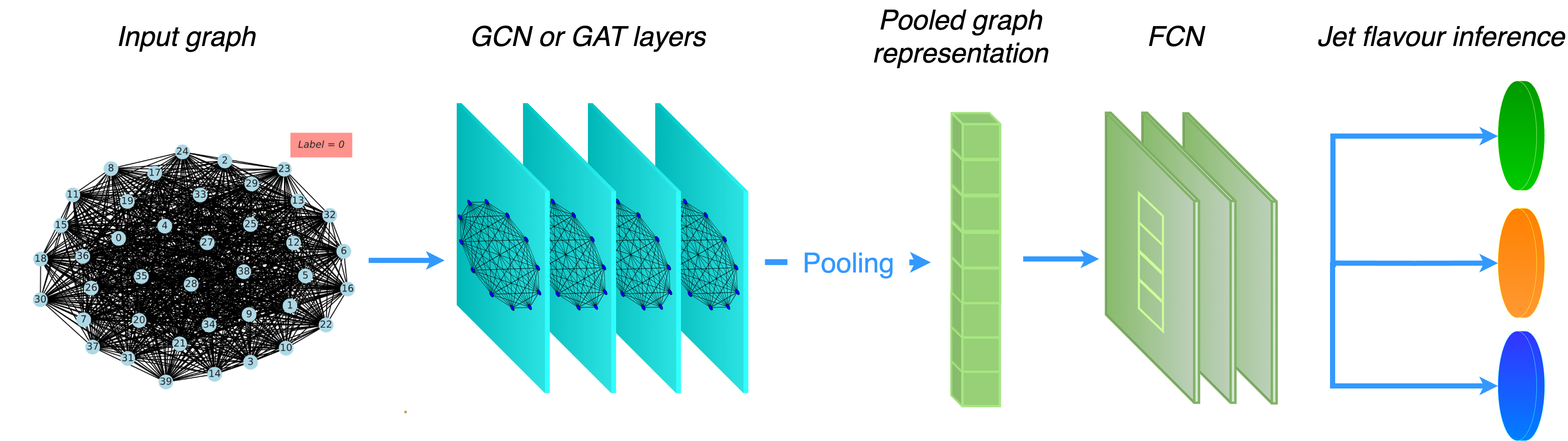}
    \caption{Representation of the network architecture. The input graph, labeled by the MC truth-level information, is classified from the Fully Connected Network in three classes corresponding to light quark, c- or b-jets. }
    \label{fig:netarchit}
\end{figure}
\noindent The network is trained by performing a classification task to identify the jets originating flavour. To feed the GNNs, the pipeline recasts the array-like dataset in a compatible format with the Graph Layers, as explained in the previous section. The graph-jets are provided as input to a user-configurable number of Graph Layers. The user is given a choice between two Message passing-based layers: the Graph Convolutional Layers (GCN) \cite{GCN} and the Graph Convolutional Attention Layers (GAT) \cite{GAT}. The configuration file allows the selection of the graph layer number and the hidden nodes per graph layer. Once the graph-jet has been processed by the GCN and/or GAT layers, a pooling function is applied. The latter aims to convert the updated graph representation into an array-like representation using embedding criteria for all track features. The user can select from various pooling functions in the configuration file. In particular, they can include a pooling attention function that applies the attention mechanism as an embedding criterion \cite{PoolATT}. A three-classes Fully Connected Network (FCN) executes the final classification, taking the pooled graph representation as input. The FCN outputs are three values representing the probability that the input jet is generated from a bottom quark, a charm quark, or a light quark. The user can set through the configuration file the number of linear layers and the number of hidden nodes per linear layer in the FCN. To sum up, the user has the ability to customize the entire architecture of the network as per their requirements. Moreover, the attention mechanism \cite{att} can be integrated into the network's graph part and the pooling function. The final architecture of the network is exemplified in Figure \ref{fig:netarchit}.

\subsection{Training and performance evaluation}
\noindent To train the previously defined network structure, the network weights need to be updated. The network adjusts the weights based on the input features and architecture to differentiate between different types of jets. The primary objective is to minimize the loss function. To achieve this, the optimizer calculates the gradient of the loss function with respect to the network weights and selects the weights that minimize the loss function. In the pipeline, the loss function and the optimizer are fixed respectively to the Cross-Entropy Loss \cite{CE} and the Adam Optimizer \cite{Adam}. The user can define the optimizer learning rate, i.e. the step size at each iteration while moving toward a minimum of a loss function. Furthermore, the network is trained on the whole training dataset in steps of data batches; the batch size can be defined by the user. \\
All the customizable hyperparameters affect the network performance and could be used to perform a grid search for network optimization.

\section{Application of the pipeline on simulated datasets}
A comprehensive testing of the pipeline to evaluate both its usability and performance has been conducted. Two different datasets have been simulated with Run 2 center-of-mass energy $\sqrt{s}=13$ TeV. The test consists of a grid search on the graph portion of the network architecture with the pooling function, the training hyperparameters and the FCN structure fixed.
\subsection{Dataset simulation}
\noindent Three Monte Carlo simulation frameworks interfaced with each other have been exploited to obtain the datasets. Firstly, MadGraph\_aMC@NLO \cite{Madgraph} was used to generate parton-level hard processes, followed by Pythia 8.3 \cite{Pythia} which provides the parton showering and hadronization. Finally, Delphes 3.5.0 \cite{Delphes} covers the detector response simulation. The ATLAS experiment card, included in Delphes by default, has been used. To extrapolate the value and the uncertainty of the transverse $d_0$ and longitudinal $z_0$ impact parameter, the Delphes configuration has been modified adding the track smearing. 

 \begin{figure}[htbp!]
\centering
		\begin{minipage}[b]{.4\textwidth}
			\centering	\includegraphics[width=1\textwidth]{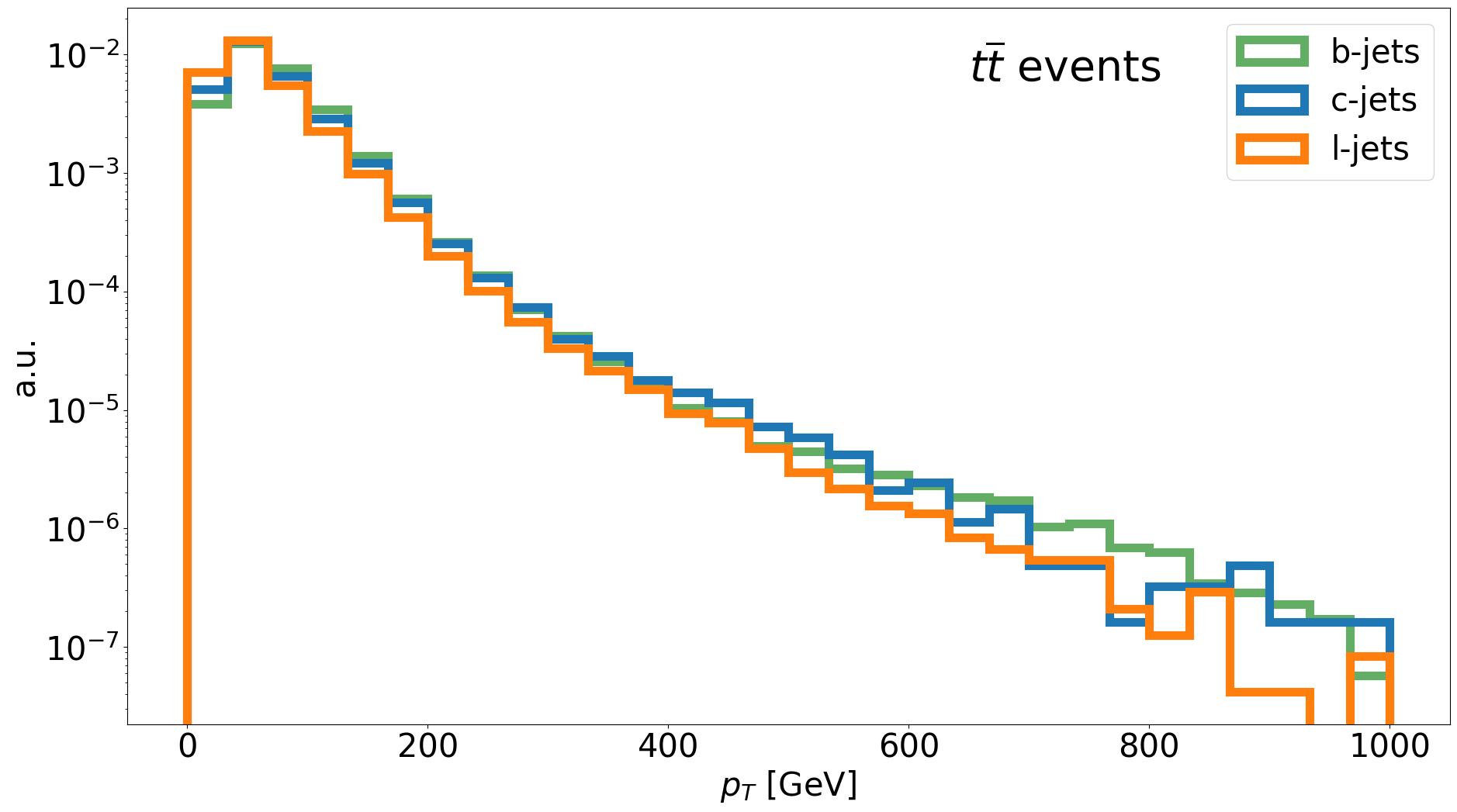}
		\end{minipage}
        \hspace{0.5cm}
		\begin{minipage}[b]{.4\textwidth}
			\centering
			\includegraphics[width=1\textwidth]{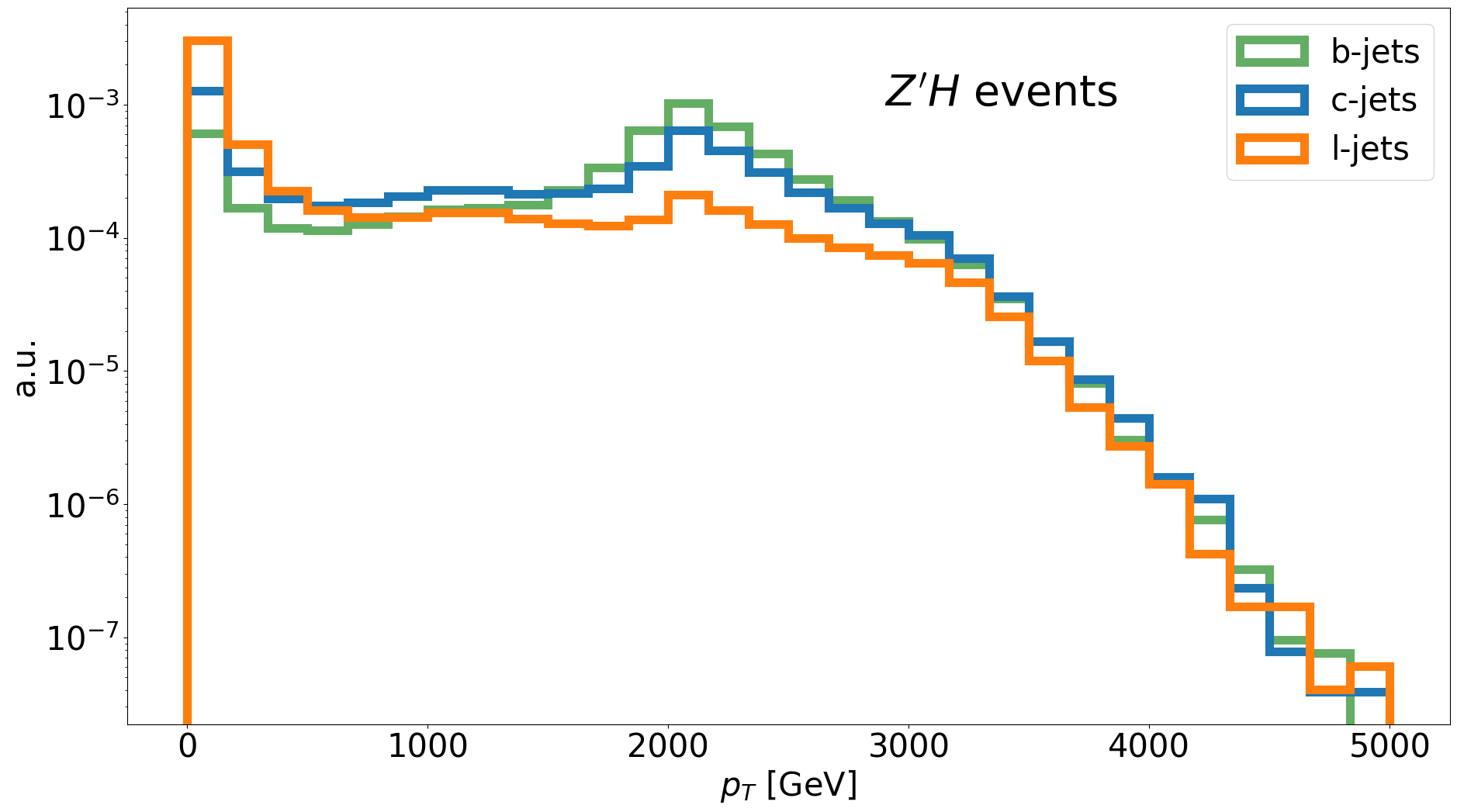}
		\end{minipage}  
  		\begin{minipage}[b]{.4\textwidth}
			\centering	\includegraphics[width=1\textwidth]{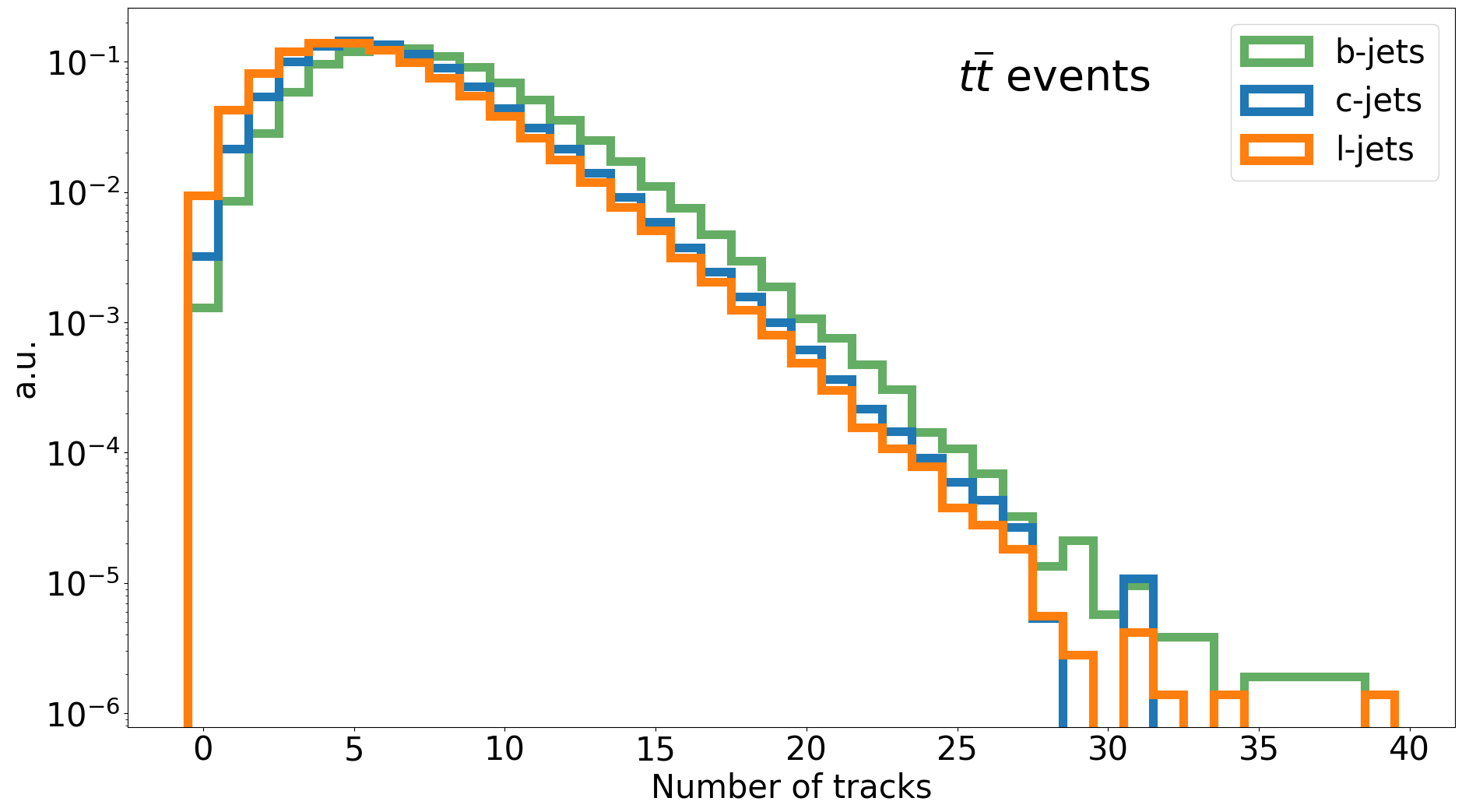}
		\end{minipage}
        \hspace{0.5cm}
		\begin{minipage}[b]{.4\textwidth}
			\centering
			\includegraphics[width=1\textwidth]{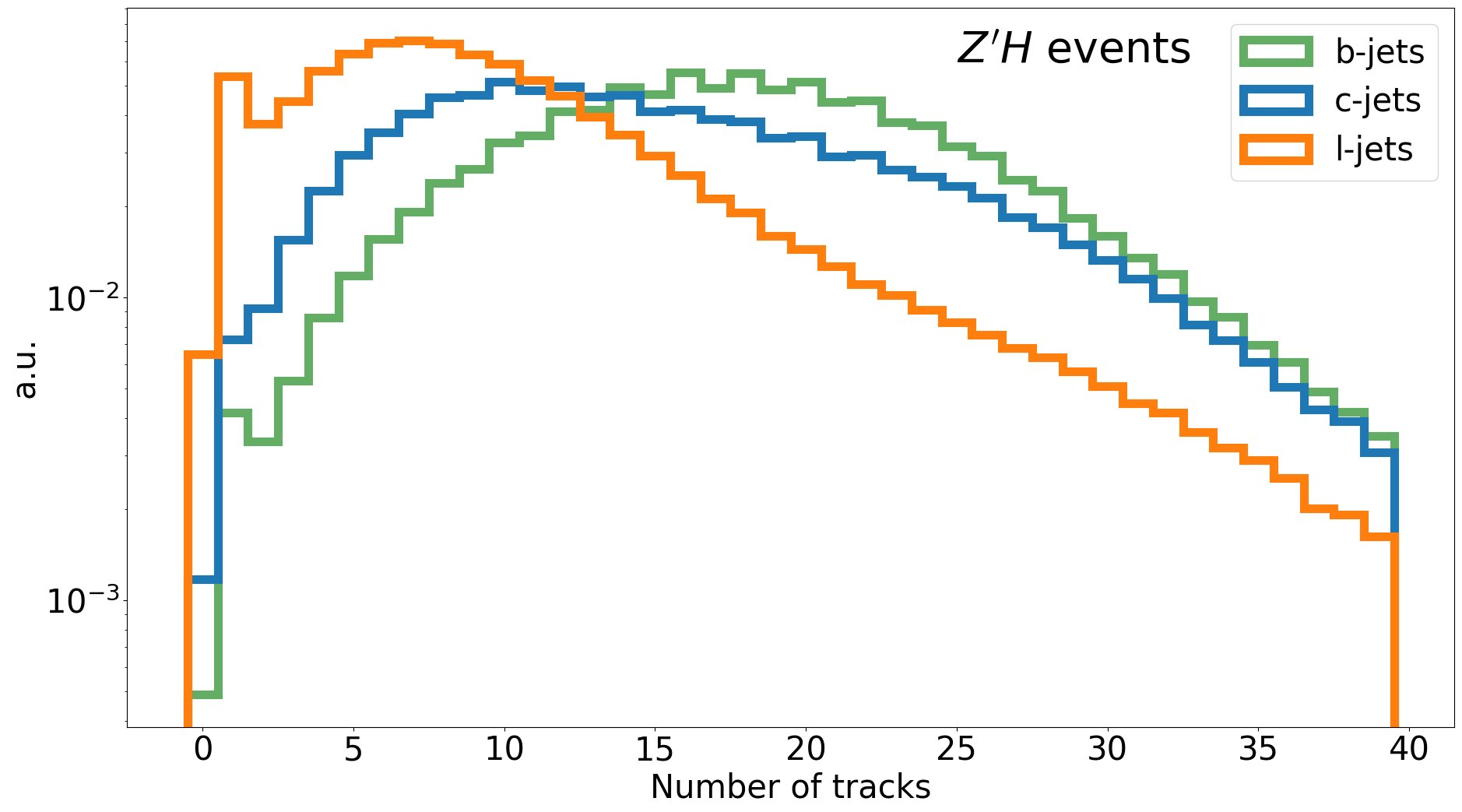}
		\end{minipage}  
\caption{Jet transverse momentum distribution for the $t\overline{t}$ at next-to-leading order dataset (up-left) and for the $Z'H$ at leading order dataset with $m_{Z'}=2$ TeV (up-right). Distributions of associated track number for the $t\overline{t}$ at next-to-leading order dataset (down-left) and for or the $Z'H$ at leading order dataset with $m_{Z'}=2$ TeV (down-right). The distributions are divided for jet truth flavours.}
\label{fig:sim_data}
\end{figure}

\noindent A next-to-leading order $t\overline{t}$ and a leading order $Z'H$ dataset have been used in the network training. In the latter, $H$ refers to the Higgs boson and $Z'$ is a hypothetical neutral vector boson that arises in extensions of the SM. Several models predict the existence of a $Z'$ boson, such as the B-L model \cite{model1} and the leptophobic $Z'$ model \cite{model2}. In this study, $Z'$ has a fixed mass of  $m_{Z'}=2$ TeV and it is restricted to decay in the hadronic channel ($Z' \rightarrow q \overline{q}$). The purpose of the $Z'H$ dataset is to extend the jet transverse momentum range as shown in the two top plots of Figure \ref{fig:sim_data}. The tracks have been associated with a jet following the $\Delta R$ criteria, where $\Delta R$ is defined as
\begin{equation}
    \Delta R = \sqrt{(\eta_{jet}-\eta_{trk})^2+(\phi_{jet}-\phi_{trk})^2}\, ,
\end{equation}
with $\eta$ the pseudorapity and $\phi$ the azimuthal angle. The track is associated with the jet whether $\Delta R \leq 0.45$ for jet $p_T \leq 150 $ GeV, while for jet $p_T > 150 $ GeV is required $\Delta R \leq 0.26$ \cite{selection}. The resulting distributions for both datasets are reported in Figure \ref{fig:sim_data}. Using the Monte Carlo truth information, the jets have been labeled with three indices 0,1,2 that respectively correspond to light, charm, and bottom jets. 
For the training, a dataset with 7 global features and 11 node features has been prepared and the description of each feature is reported in Table \ref{tab:my_label}.
\begin{table}[htbp!]
    \centering
    \begin{tabular}{|p{2cm}|p{13cm}|}
    \hline
        \textbf{Jet Input} & \textbf{Description}\\
        \hline
         $p_T$ &  Transverse momentum \\
         $\eta$ &  Pseudorapidity\\
         $\phi$ &  Azimuthal angle\\
         $\Delta \eta$ & Difference between pseudorapidities of the two most energetic particles in the jet \\
         $\Delta \phi$ & Difference between azimuthal angles of the two most energetic particles in the jet \\
         $E_h/E_e$ & Ratio of the hadronic versus electromagnetic energy deposited in the calorimeter \\
         $N_{trk}$ &  Number of associated tracks\\
        \hline
        \hline
        \textbf{Track Input} & \textbf{Description}\\
        \hline
         $p_T$ &  Transverse momentum \\
         $q/P$ &  Track charge divided by momentum (measure of curvature) \\
         $\delta \eta$ &  Pseudorapidity of the track, relative to the jet $\eta$\\
         $\delta \phi$ &  Azimuthal angle of the track, relative to the jet $\phi$\\
         $z_0 sin(\theta)$ & Closest distance from the track to the PV in the longitudinal plane \\
         $d_0$ &  Closest distance from the track to the PV in the transverse plane\\
         $\sigma(q/P)$ & Uncertainty on $q/P$\\
         $\sigma(\theta)$ & Uncertainty on polar angle $\theta$\\ 
         $\sigma(\phi)$ & Uncertainty on azimuthal angle $\phi$\\
         $s(z_0)$ & Lifetime signed longitudinal IP significance \\
         $s(d_0)$ & Lifetime signed transverse IP significance \\
    \hline
    \end{tabular}
    \caption{Input features to the network. Basic jet kinematics, along with information about the reconstructed track parameters are used.}
    \label{tab:my_label}
\end{table}

\subsection{Results}
\begin{figure}[htbp!]
     \centering
		\begin{minipage}[b]{.3\textwidth}
			\centering	
   \includegraphics[width=1\textwidth]{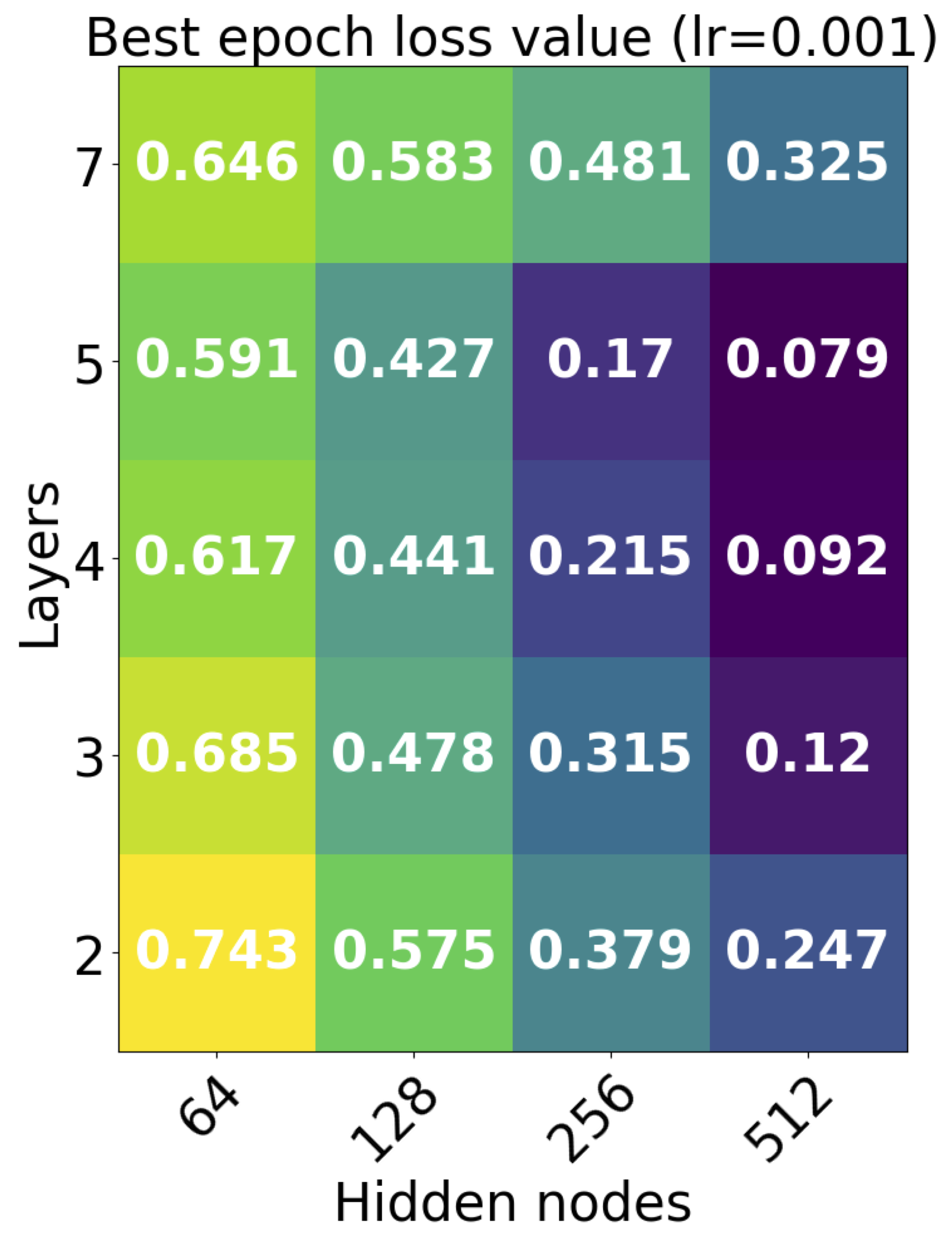}
		\end{minipage}
        \hspace{0.3cm}
		\begin{minipage}[b]{.3\textwidth}
			\centering
			\includegraphics[width=1\textwidth]{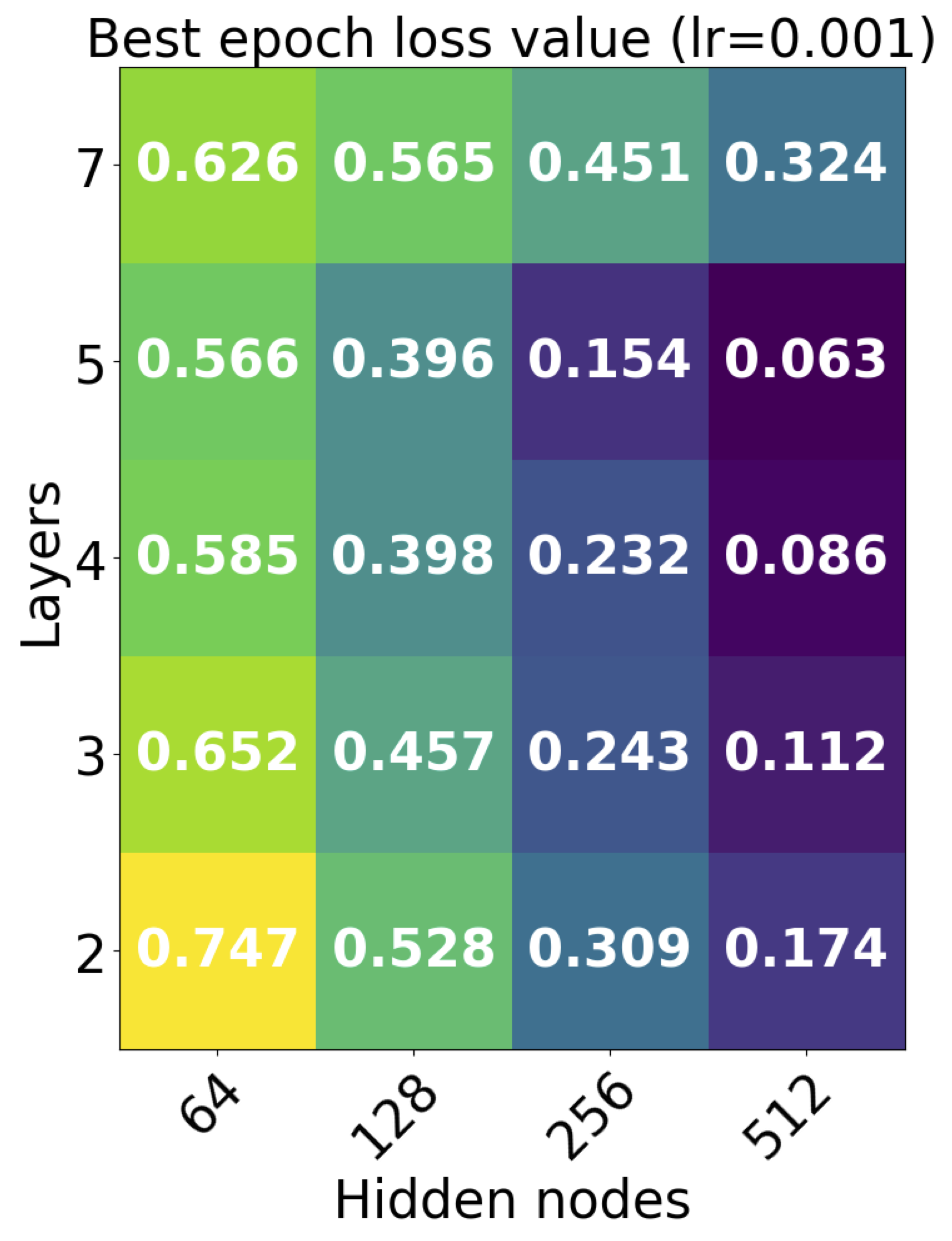}
		\end{minipage}  
  \caption{Results from the grid search. (left) Heat map with best epoch loss value for the grid search on $t\overline{t}$ at the next-to-leading order dataset (right). Heat map with best epoch loss value for the grid search on $Z'H$ at the leading order dataset. }
  \label{fig:loss}
\end{figure}

\begin{figure}[htbp!]
     \centering
		\begin{minipage}[b]{.6\textwidth}
			\centering	
   \includegraphics[width=1\textwidth]{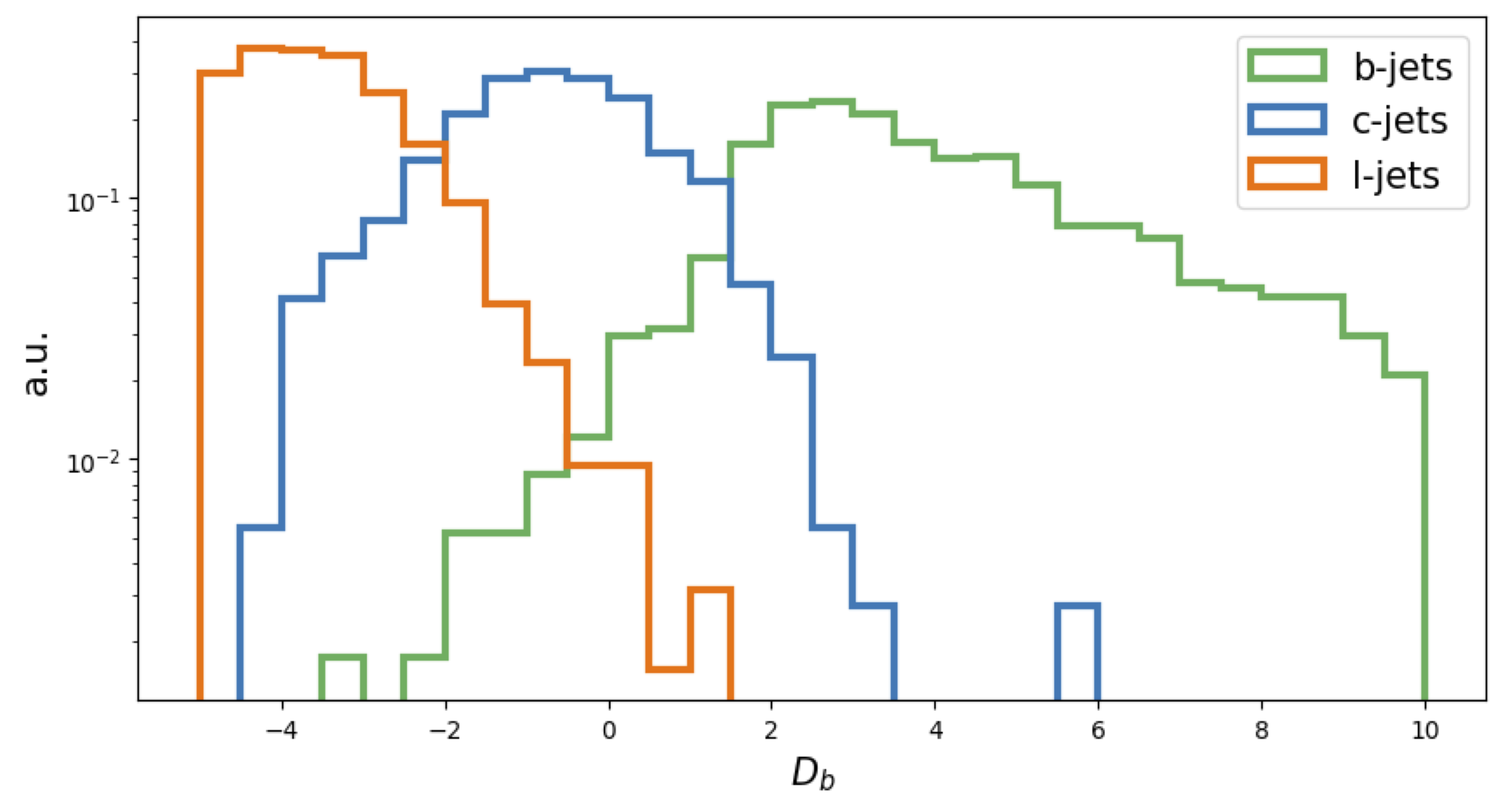}
		\end{minipage}
        \hspace{0.3cm}
		\begin{minipage}[b]{.7\textwidth}
			\centering
			\includegraphics[width=1\textwidth]{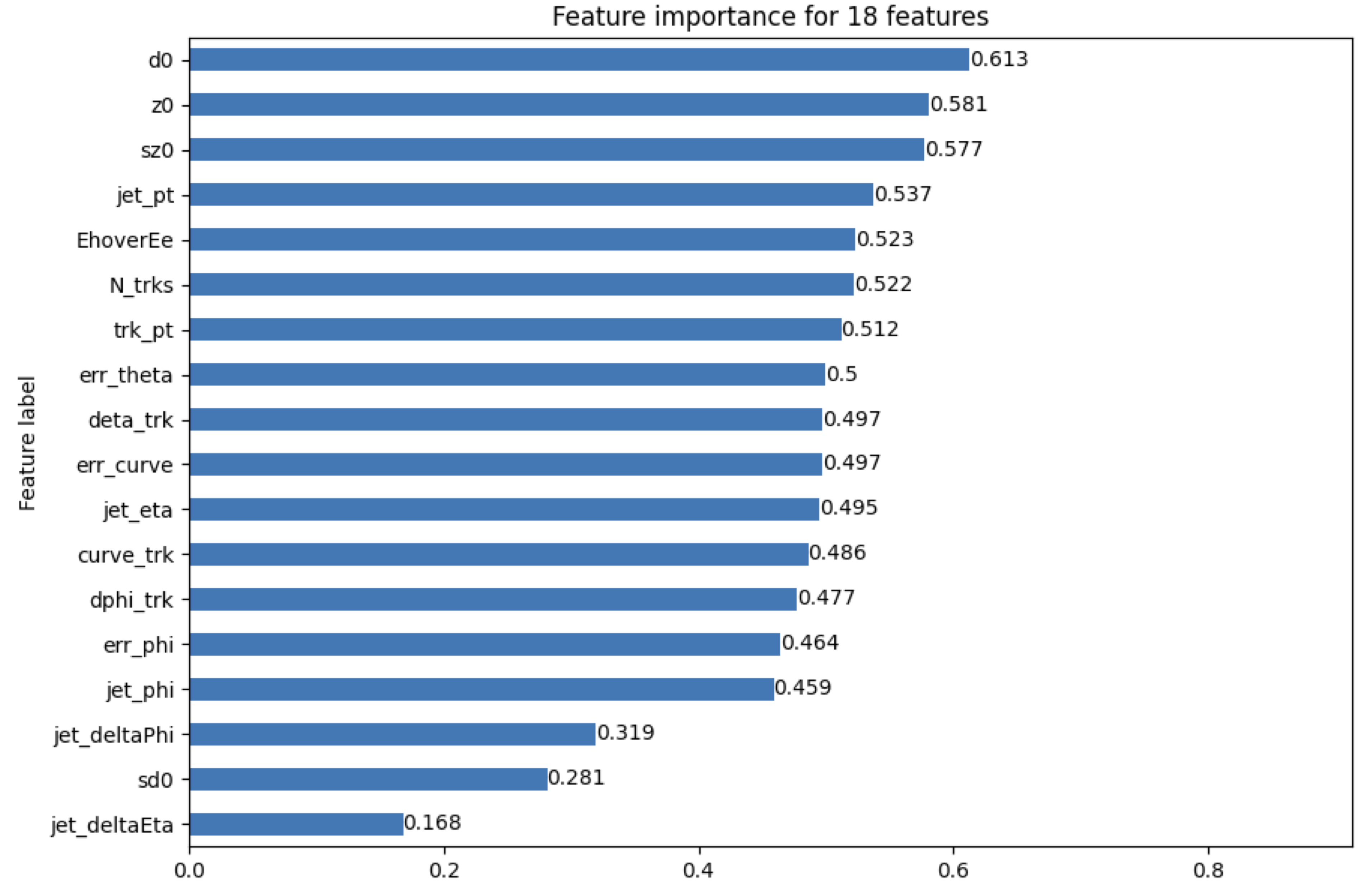}
		\end{minipage}  
  \caption{Discriminant distribution for the best architecture with the $t\overline{t}$ at leading order dataset divided per truth flavour (up). Representation of the 18 input features ranking (down).}
  \label{fig:resss}
\end{figure}

\begin{table}[htbp!]
    \centering
    \begin{tabular}{|p{5cm}|p{5cm}|}
    \hline
        \textbf{Hyperparameter} & \textbf{Value}\\
        \hline
        Epochs & 500 \\
        GCN layers & [2,3,4,5,7]\\
        GCN hidden nodes & [64,128,256,512]\\
        FCN linear layers & 3\\
        FCN hidden nodes & 64 \\
        Pooling function & global\_mean\_pool \\
        Loss function & Cross Entropy Loss \\
        Optimizer & Adam \\
        Learning rate & 0.001 \\
        \hline
    \end{tabular}
    \caption{Hyperparameters used in the grid search performed on 20 different architectures.}
    \label{tab:hyper}
\end{table}
We conducted a grid search using both datasets. In this search, we examined a total of 20 different architectures, each trained for 500 epochs with the hyperparameters listed in Table \ref{tab:hyper}. This approach allowed us to explore the various architectures to identify the best-performing model. To analyze the performance of the different architectures tested, we have extracted the minimum value assumed by the loss function during the training process over the 500 epochs. This metric provides insights into how well the model is able to minimize the difference between predicted and actual values. The obtained minimum loss value for each architecture has been plotted in Figure \ref{fig:loss} for both datasets. By examining the plot, we can compare the performance of the various architectures and determine which one achieves the lowest loss value.
\noindent The results of our analysis indicate that the architecture with 5 GCN layers and 512 hidden nodes per GCN layer has the lowest value of the loss function for both datasets. This suggests that this architecture is the most optimal for our purposes. However, we acknowledge that evaluating the network's performance based solely on the loss function may not be sufficient. Therefore, the pipeline provides the distribution of the discriminant, $D_b$, which can offer additional insights into the network's performance. That discriminant is defined as
\begin{equation}
    D_b = log\left(\frac{p_b}{p_c f_c + (1-f_c) p_l}\right) \, , 
\end{equation} 
where $p_b$, $p_c$ and $p_l$ are the network output probabilities for b-,c- and light quarks respectively, and $f_c$ is the fraction of c-jet in the dataset. This variable represents the network's capability to distinguish between jet flavours. By examining the $D_b$ distribution, we can better understand how well the network can discriminate between different classes, and make more informed decisions about its efficacy. The discriminant distribution obtained training the architecture with 5 GCN layers and 512 hidden nodes per GCN layer on the $t\overline{t}$ dataset is shown in Figure \ref{fig:resss} (up).
\noindent In order to gain a better understanding of events, it is essential to extract physical information using tagging algorithms. The pipeline facilitates this by providing a feature ranking of the input features. By ranking these features based on their importance in the network classification task, we can gain insights into the physical properties of the events. Currently, the Pytorch Geometric Explainer method \cite{Explainer} is used in the pipeline to rank these features. This method provides a comprehensive approach to analyze and evaluate features that are crucial to the classification of events. However, we are continuously working on improving this feature ranking process to ensure that it is accurate and efficient. The result of the feature ranking obtained with the trained model with which we produced the discriminant distribution is shown in Figure \ref{fig:resss} (down). From the latter, it is evident that the network considers the impact parameters to be fundamental, as they provide crucial information for determining the characteristics of the jet being analyzed.

\section{Conclusion}
\noindent The AUTOGRAPH pipeline, a state-of-the-art flavour tagging algorithm based on GNNs, has been presented. The pipeline consists of several components that work together to provide accurate and efficient jet tagging of different flavours produced in high-energy physics experiments. Firstly, the input data is pre-processed to extract relevant features that are then fed into the GNN-based tagging algorithm. The GNN architecture comprises several layers of neural networks that learn to recognize patterns in the input data. The output of the GNN is a set of probabilities that correspond to each flavour of the particle. The AUTOGRAPH pipeline has been extensively tested on two simulated datasets to evaluate its usability and performance. The performance of the pipeline was evaluated using standard metrics. Overall, the AUTOGRAPH pipeline provides a straightforward and efficient way to perform flavour tagging in high-energy physics experiments.

\end{document}